\documentclass[sn-mathphys-num,iicol]{sn-jnl}
\usepackage{graphicx}
\usepackage{bm}
\usepackage{amssymb,amsmath,latexsym}
\usepackage{xcolor}
\usepackage{array}
\usepackage{slashed}

\DeclareMathOperator{\Tr}{Tr}

\begin{document}

\title{Describing the speed of sound peak of isospin-asymmetric cold strongly interacting matter using effective models}

\author[1]{\fnm{Alejandro} \sur{Ayala}}\email{ayala@nucleares.unam.mx}

\author[2]{\fnm{Bruno} \sur{S. Lopes}}\email{bruno.lopes@acad.ufsm.br}

\author*[2]{\fnm{Ricardo} \sur{L. S. Farias}}\email{ricardo.farias@ufsm.br}

\author[1]{\fnm{Luis} \sur{C. Parra}}\email{luis.parra@correo.nucleares.unam.mx}

\affil[1]{\orgdiv{Instituto de Ciencias Nucleares}, \orgname{Universidad Nacional Aut\'onoma de M\'exico}, \orgaddress{\street{Circuito Ext. S/N Ciudad Universitaria}, \postcode{04510 CdMx}, \country{Mexico}}}

\affil[2]{\orgdiv{Departamento de F\'isica}, \orgname{Universidade Federal de Santa Maria}, \orgaddress{\city{Santa Maria}, \postcode{RS 97105-900}, \country{Brazil}}}


\abstract{The non-monotonic behavior of the speed of sound for isospin imbalanced strongly interacting matter, found by recent lattice QCD simulations, can be reproduced within the Nambu--Jona-Lasinio model and Linear Sigma Model with quarks when the couplings become isospin chemical potential-dependent. The introduction of medium-dependent couplings can potentially affect the equivalence between the thermodynamic relations and their definitions from statistical mechanics. We describe the procedure to compensate for the introduction of medium-dependent couplings to preserve the correct thermodynamic identities. We find the isospin chemical potential dependence for the couplings from the isospin density LQCD data and, after finding the compensating function to correctly describe the pressure, we show that the description of the square of the speed of sound reported by LQCD is well reproduced when using the found medium-dependent couplings in both models.}


\maketitle

\section{Introduction}
It is by now widely accepted that Quantum Chromodynamics (QCD) possesses a rich phase structure that can be revealed when strongly interacting systems are subject to varying external conditions such as temperature and density or in the presence of electromagnetic fields (for recent reviews on the subject see for example~\cite{Fukushima:2010bq,Bali:2011qj,Kharzeev:2013jha,Miransky_2015}). Of foremost importance are the QCD properties under extreme conditions encoded in the temperature ($T$) vs. baryon chemical potential ($\mu_B$) phase diagram, that can be experimentally studied by means of relativistic heavy-ion collisions, when varying the collision energy~\cite{Agarwal:2022ydl,MPD:2022qhn,Zbroszczyk:2022qpf}. Other important aspects of strongly interacting systems can be studied in the case where there is an imbalance in the isospin charge, characterized by an isospin chemical potential $\mu_I$~\cite{Brandt:2017oyy,Brandt:2022hwy,Aryal:2020ocm}. Nature provides us with systems that are subject to such conditions in the form of dense astrophysical objects such as neutron stars~\cite{Lattimer:2000nx,Lattimer:2004pg,Lattimer:2006xb}.

From the theoretical point of view, first principles calculations to find the phase structure in the $T$-$\mu_B$ plane are hindered by the notorious sign problem~\cite{Karsch:2001cy,Muroya:2003qs}. This problem has been circumvented by resorting to calculations that use effective models implementing the symmetries of QCD at low energies~\cite{Gutierrez:2013sta,Roessner:2006xn,Ayala:2019skg,Asakawa:1989bq,Ayala:2017ucc,Gao:2020fbl}. Nevertheless, Lattice QCD (LQCD) calculations for finite $\mu_I$ do not suffer from the sign problem and thus represent useful test grounds for these effective models~\cite{Son:2005qx,Son:2000xc,Son:2000by,Splittorff:2000mm,Loewe:2002tw,Loewe:2005yn,Fraga:2008be,Cohen:2015soa,Janssen:2015lda,Carignano:2016lxe,Lepori:2019vec,Adhikari:2020kdn,Adhikari:2020qda,Adhikari:2020ufo,ADHIKARI2020135352,PhysRevD.106.114017,Adhikari:2019mdk,Gao:2016qkh,Andersen:2015eoa,Frank:2003ve,Toublan:2003tt,Barducci:2004tt,He:2005sp,He:2005nk,He:2006tn,Ebert:2005cs,Ebert:2005wr,Sun:2007fc,Andersen:2007qv,Abuki:2008wm,Xia:2013caa,Khunjua:2018jmn,Khunjua:2018sro,Khunjua:2017khh,Ebert:2016hkd,Mukherjee:2006hq,Bhattacharyya:2012up,Kamikado:2012bt,Ueda:2013sia,
Stiele:2013pma,Adhikari:2018cea,Braun:2022olp}. 

An important result obtained by recent LQCD calculations is the existence of a peak in the square of the speed of sound ($c_s^2$) at intermediate values of $\mu_I$. The square of the speed of sound exhibits a non-monotonic behavior, reaching a maximum above the conformal limit $c_s^2=1/3$ to then decrease toward that limit from above~\cite{Brandt:2022hwy,Brandt:2022fij}. Although the calculation uncertainties increase as $\mu_I$ grows this result is nowadays regarded as a messenger of important properties of strongly interacting matter. For instance, in Ref.~\cite{Chiba:2023ftg} the peak in the square of the speed of sound is interpreted  as the signature of the quark saturation for the onset of quark matter formation. A similar idea has been explored in Ref.~\cite{Cao:2020byn} in the context of the emergence of quarkyonic matter with increasing density, or the transition from hadron to quark dominated matter also with increasing density in Refs.~\cite{Kojo:2021hqh,Kojo:2022psi}, although in Ref.~\cite{Mu:2010zz} it has also been argued that charged pions are still bound states even for large values of $\mu_I$. More recently, the large isospin chemical potential region of the QCD phase diagram has been explored on the lattice~\cite{Abbott:2023coj} and a peak in the energy density was identified for $\mu_I \sim 1.5 \ m_\pi$, which is argued to signal the transition into a Bose-Einstein condensed phase.

In this work we put forward the idea that the peak in the speed of sound can be described by introducing $\mu_I$-dependent couplings. That couplings may depend on the properties of the medium is not surprising. In thermo-magnetic QCD it is known that the coupling $\alpha_s$ runs with the temperature and field strength~\cite{Ayala:2018wux}. In the context of effective models, this idea has been explored to describe inverse magnetic catalysis using the Nambu--Jona-Lasinio (NJL) model~\cite{Farias:2014eca,Farias:2016gmy,Avancini:2016fgq,Tavares:2021fik,Avancini:2018svs} and the Linear Sigma Model with quarks (LSMq)~\cite{Ayala:2014iba,Ayala:2021nhx}. In the former case, a functional form for the $T$ and $B$-dependent coupling is introduced to mimic asymptotic freedom. In the latter, the $T$ and $B$-dependent couplings are computed from the model itself at one-loop order. In this work we try a different but related approach. We find the $\mu_I$ dependence of couplings that describe the isospin density which carries the most direct information of the properties of the speed of sound, utilizing as inputs the lattice results from Refs.~\cite{Brandt:2022hwy,Brandt:2022fij}. However, the introduction of a medium dependent parameter to the system via the coupling has consequences for the thermodynamic identities. In this particular case, the isospin density, as computed by its statistical mechanical definition, would not coincide with that computed from the thermodynamic relations. Under the assumption that these calculations should coincide, a compensating function can be introduced to the system's Hamiltonian (and consequently to the effective pressure) and be subsequently determined by identifying the coupling variations. The subject of the thermodynamic consistency of systems with medium dependent parameters has been explored in the literature to a considerable extent. In Ref.~\cite{Gorenstein:1995vm}, the authors generalized the main ideas within a study of a phenomenological model with a gluon medium dependent mass. Further applications in a wide range of contexts may be found in Refs.~\cite{Schertler:1996tq,Levai:1997yx,Wang:2000dc,Yin:2008me,Lenzi:2010mz,Xia:2014zaa,Restrepo:2022wqn,Ma:2023stj}. Here, we explicitly explore this scenario using our two-flavor NJL and LSMq descriptions of isospin imbalanced matter at $T=0$~\cite{Avancini:2019ego,Lopes:2021tro,Ayala:2023cnt}. Notice that the LQCD results, used in this work to compare with our model calculations, are performed using 2+1 flavors. In the LQCD calculation, the strange quark is included at the action level with zero chemical potential ($\mu_s = 0$). While the operator $n_I$ depends only on $\mu_u$ and $\mu_d$, the expectation value of the isospin density, $\left\langle n_I \right\rangle$, also depends on $\mu_s$ via the determinant weights in the path integral.  This means that in a perturbative calculation, the effect of the strange quark on $\left\langle n_I \right\rangle$ can  only be captured at higher loop orders involving $s$-quark bubbles. In the spirit of a perturbative calculation, these higher order contributions are thus suppressed with respect to a one-loop order calculation, such as what we present in this work. Therefore a comparison between our two-flavor model analysis and the referenced LQCD data is justified within physical grounds since our model calculations capture the leading order effect. Moreover, it has been observed in Refs.~\cite{ADHIKARI2020135352,Lopes:2021tro} that the two-flavor NJL model and chiral perturbation theory calculations are in better agreement with LQCD results in comparison with their respective three-flavor versions.

The work is organized as follows: In Sec.~\ref{secII} we discuss in general the subject of thermodynamic consistency when a parameter that is originally considered as medium-independent is effectively described as medium-dependent. In Secs.~\ref{secIII} and~\ref{secIV}, we introduce the NJL and LSMq models, respectively, and make the couplings become medium-dependent. We fit the isospin density to find the explicit $\mu_I$ dependence of the couplings, and apply the thermodynamic consistency procedure to find the description of the rest of the thermodynamic quantities. We show that this procedure provides a good description in particular of the peak structure of the square of the speed of sound as a function of $\mu_I$. We finally summarize and discuss our results in Sec.~\ref{concl}.
%
\section{Thermodynamic consistency}\label{secII}
The subject of the thermodynamically consistent description of a physical system, when introducing effective medium dependent quantities, has been addressed in the literature in different contexts~\cite{Gorenstein:1995vm,Schertler:1996tq,Levai:1997yx,Wang:2000dc,Yin:2008me,Lenzi:2010mz,Xia:2014zaa,Restrepo:2022wqn,Ma:2023stj}. For our purposes, we discuss the subject when introducing a medium-dependent coupling. Under these circumstances, the Hamiltonian, which is originally taken as being medium-independent, now becomes medium-dependent. This leads to an apparent contradiction between the usual thermodynamic relations and their definition using statistical mechanics. To see this, consider the statistical mechanics definitions of the pressure $P$, energy density $\varepsilon$ and number density $n$ for a grand-canonical ensemble described by the inverse temperature $\beta$ and a chemical potential $\mu$
\begin{eqnarray}
P(T,\mu) &=& \frac{T}{V} \ln \Tr \left[e^{-\beta(H - \mu N)}\right] , 
\label{eqn:stat_mec_defs-1}\\
\varepsilon(T,\mu) &=& \frac{1}{V} \frac{1}{Z(T,\mu,V)} \Tr \left[H e^{-\beta(H - \mu N)}\right] , 
\label{eqn:stat_mec_defs-2}\\
n(T,\mu) &=& \frac{1}{V} \frac{1}{Z(T,\mu,V)} \Tr \left[N e^{-\beta(H - \mu N)} \right] ,
\label{eqn:stat_mec_defs}
\end{eqnarray}
where $Z(T,\mu,V)$ is the partition function, $H$ is the Hamiltonian and $N$ is a conserved charge number operator. When in a model represented by $H$ a given parameter, say a coupling constant $G$, depends on the chemical potential, namely $G=G(\mu)$, then $H\to H(\mu)$ and the number density, obtained by means of the thermodynamic relation
\begin{eqnarray}
n = \left(\frac{\partial P}{\partial \mu}\right)_T \, ,
\label{eqn:thermo_ids}
\end{eqnarray}
no longer coincides with Eq.~\eqref{eqn:stat_mec_defs}. This apparent ambiguity in the definition of the number density may then have consequences for the energy density $\varepsilon$
\begin{eqnarray}
\varepsilon &=& - P + Ts + \mu n \, ,
\end{eqnarray}
where $s = \left(\partial P / \partial T\right)_\mu$ is the entropy density. A self-consistent solution for this problem is to consider an additional contribution to the Hamiltonian~\cite{Gorenstein:1995vm}, producing an effective Hamiltonian $H_{\textrm{eff}}$. We require that Eqs.~\eqref{eqn:stat_mec_defs-1} -- \eqref{eqn:stat_mec_defs} are satisfied by $H_{\textrm{eff}} = H + E^*$, where $E^*$ also depends on the medium through the model parameter $G(\mu)$, such that $H_{\textrm{eff}}$ does not have an explicit $\mu$-dependence, and thus that
\begin{align}\label{eqn:constraint_hamiltonian}
\left(\frac{\partial E^*}{\partial \mu}\right)_T &= - \left(\frac{\partial H}{\partial \mu}\right)_T \, , \nonumber \\
&= - \left(\frac{\partial H}{\partial G}\right)_{T,\mu} \left(\frac{\partial G}{\partial \mu}\right)_T \, .
\end{align}
The approach is thermodynamically consistent provided this constraint is satisfied. The pressure, however, will change due to the energy shift in the Hamiltonian. Letting $B^* = E^* / V$, the effective pressure $P_{\textrm{eff}}$ can be written as
\begin{align}\label{eqn:pres_tilde}
P_{\textrm{eff}} = P - B^* \, ,
\end{align}
where $P$ is given by the usual expression evaluated with $G = G(\mu)$. To find $B^*$ we notice that
\begin{align}
\label{eqn:partial}
\left(\frac{\partial P_\textrm{eff}}{\partial \mu}\right)_T = \left(\frac{\partial P_\textrm{eff}}{\partial \mu}\right)_{T,G} + 
\left(\frac{\partial P_\textrm{eff}}{\partial G}\right)_{T,\mu} \left(\frac{\partial G}{\partial \mu}\right)_T \, .
\end{align}
The first term on the right-hand side yields exactly the density as defined from Eq.~\eqref{eqn:stat_mec_defs}, but for the effective Hamiltonian. Thus, for the number density to keep its form, we may alternatively require that 
\begin{align}
\left(\frac{\partial P_{\textrm{eff}}}{\partial G}\right)_{T,\mu} = 0 \, ,
\end{align}
which is equivalent to the constraint in Eq.~\eqref{eqn:constraint_hamiltonian}. From this condition and the definition from Eq.~\eqref{eqn:pres_tilde}, we get by integration
\begin{align}\label{eqn:def_bstar}
B^*(\mu) = B_0^* + \int_{\mu_0}^{\mu} \left(\frac{\partial P}{\partial G}\right)_{T,\mu} \left(\frac{\partial G}{\partial \mu^\prime}\right)_T d\mu^\prime \, ,
\end{align}
where $B_0^*$ is an integration constant and $\mu_0$ can be taken as the chemical potential corresponding to the onset of a finite number density. Since the thermodynamics is not affected by a constant, hereafter we ignore $B_0^*$.

For our purposes, we work at $T=0$ but with $\mu_I\neq 0$. The model pressure is either $P_{\textrm{NJL}}$ or $P_{\textrm{LSMq}}$ which play the role of $P$ in the previous expressions. The dependence of $G$ on $\mu_I$ is obtained by fitting the model to the LQCD data~\cite{Brandt:2022fij,Brandt:2022hwy} for the isospin density to find the $G(\mu_I)$ that best describes the data. In the following sections, we aim to show how this running of the couplings plays a crucial role in the description of the non-monotonic behavior found for the speed of sound.
%
\section{The NJL model at finite isospin chemical potential}\label{secIII}
The formalism of the two-flavor NJL model at finite isospin density has been a subject of considerable scrutiny in the literature~\cite{Toublan:2003tt,Frank:2003ve,Barducci:2004tt,He:2005sp,He:2005nk,He:2006tn,Ebert:2005wr,Ebert:2005cs,Andersen:2007qv,Sun:2007fc,Abuki:2008wm,Mu:2010zz,Xia:2013caa,Ebert:2016hkd,Khunjua:2017khh,Khunjua:2018sro,Khunjua:2018jmn,Khunjua:2019lbv,Khunjua:2019ini,Avancini:2019ego,Lu:2019diy,Khunjua:2020xws,Lopes:2021tro,Khunjua:2021oxf,Liu:2021gsi}. Here, some of the key elements are revisited. We start by writing the partition function at finite temperature $T$, baryon chemical potential $\mu_B$ and isospin chemical potential $\mu_I$
\begin{align}
&Z_{\textrm{NJL}} (T,\mu_B,\mu_I) = \int \left[d\bar{\psi}\right]\left[d\psi\right] \times \nonumber \\
&\exp \left[ - \int_0^\beta d\tau \int d^3 x \left( \mathcal{L}_{\textrm{NJL}} + \bar{\psi}\hat{\mu}\gamma_0\psi \right) \right] \, ,
\end{align}
in which $\hat{\mu} = \textrm{diag}\left(\mu_u,\mu_d\right)$ is the quark chemical potential matrix in flavor space, with $\mu_{u,d}$ of the individual quark flavors given by
\begin{align}
&\mu_u = \frac{\mu_B}{3} + \frac{\mu_I}{2} \, , \nonumber \\
&\mu_d = \frac{\mu_B}{3} - \frac{\mu_I}{2} \, ,
\label{choice1}
\end{align}
and, consequently, we can write $\mu_B / 3 = (\mu_u + \mu_d) / 2$ and $\mu_I = \mu_u - \mu_d$. In what follows, we consider only the case $\mu_B=0$. The Lagrangian density of the model can be written as
\begin{align}\label{eqn:njl_lagrangian}
\mathcal{L}_{\textrm{NJL}} &= \bar{\psi}\left(i\slashed{\partial} - m\right)\psi + G \left[ \left(\bar{\psi}\psi\right)^2 + \left(\bar{\psi}i\gamma_5 \vec{\tau} \psi\right)^2 \right] \nonumber \\
&= \bar{\psi}\left(i\slashed{\partial} - m\right)\psi + G \left[ \left(\bar{\psi}\psi\right)^2 + \left(\bar{\psi}i\gamma_5 \tau_3 \psi\right)^2 \right. \nonumber \\
&\left.+ 2 \left(\bar{\psi}i\gamma_5 \tau_{+}\psi\right)\left(\bar{\psi}i\gamma_5 \tau_{-}\psi\right)\right] \, ,
\end{align}
where $\psi = (u,d)^T$ represents the quark fields and $m$ is the current quark mass. The scalar/pseudoscalar coupling is denoted by $G$ and $\vec{\tau} = (\tau_1,\tau_2,\tau_3)$ are the Pauli matrices, namely the $SU(2)$ group generators. Alternatively, we can utilize the representation ($\tau_3,\tau_+,\tau_-$), defining the combinations $\tau_{\pm} = (\tau_1 \pm i\tau_2)/\sqrt{2}$, such that a correspondence to the neutral ($\pi_3$) and charged pion excitations ($\pi_\pm$) can be made.
\par The presence of a finite isospin chemical potential $\mu_I$ explicitly breaks the isospin symmetry group $SU(2)$ down to a subgroup $U(1)_{I_3}$, with the third component of the isospin charge $\mathbf{I_3}$ being the generator~\cite{Mu:2010zz}. Thus, in the context of the mean-field approximation, we can consider $\left\langle \bar{\psi}i\gamma_5 \tau_3 \psi \right\rangle = 0$ as an ansatz, while allowing the possibility of $\left\langle \bar{u}i\gamma_5 d \right\rangle = \left\langle \bar{d}i\gamma_5 u \right\rangle^{*} \neq 0$. The chiral condensate $\sigma$ and pion condensate $\Delta$ are introduced as
\begin{align}
\sigma &= - 2 G \left\langle \bar{\psi} \psi \right\rangle \, , \\
\sqrt{2}\pi_+ &= - 2\sqrt{2} G \left\langle \bar{\psi} i\gamma_5 \tau_+ \psi \right\rangle = \Delta e^{i \theta} \, , \\
\sqrt{2}\pi_- &= - 2\sqrt{2} G \left\langle \bar{\psi} i\gamma_5 \tau_- \psi \right\rangle = \Delta e^{-i \theta} \, ,
\end{align}
with the phase factor $\theta$ indicating the direction of the $U(1)_{I_3}$ symmetry breaking. We adopt $\theta = 0$, a choice equivalent to considering only $\left\langle \bar{\psi}i\gamma_5 \tau_1 \psi \right\rangle \neq 0$ as any nonzero value of $\left\langle \bar{\psi}i\gamma_5 \tau_2 \psi \right\rangle$ can be rotated away~\cite{Andersen:2007qv}.
\par In the mean-field approximation, the thermodynamic potential is given by
\begin{align}
\Omega_{\textrm{NJL}} = \frac{\sigma^2 + \Delta^2}{4 G} - 2 N_c \int_\Lambda \frac{d^3 k}{(2\pi)^3}\left(E_{k}^{+} + E_{k}^{-}\right) \, ,
\label{eqn:njl_omega}
\end{align}
where $E_{k}^{\pm} = \sqrt{\left(E_k \pm \frac{\mu_I}{2} \right)^2 + \Delta^2}$, $E_k = \sqrt{k^2 + M^2}$, and $M = m + \sigma$ is the effective mass. The symbol $\int_\Lambda$ indicates integrals that need to be regularized, and we utilize the method of a sharp ultraviolet momentum cutoff $\Lambda$~\cite{Klevansky:1992qe}. The physical values of the condensates $\sigma$ and $\Delta$ are determined by minimizing the thermodynamic potential, i.e. by solving the gap equations
\begin{align}\label{eqn:njl_gap_eqs}
\left.\frac{\partial \Omega_{\textrm{NJL}}}{\partial \sigma}\right|_{\sigma = \sigma_m} = \left.\frac{\partial \Omega_{\textrm{NJL}}}{\partial \Delta}\right|_{\Delta = \Delta_m} = 0 \, .
\end{align}
\par With $\Omega_\textrm{NJL}$ defined, we can obtain the thermodynamic quantities of interest. The isospin density $n_I = - \partial \Omega_\textrm{NJL} / \partial \mu_I$ is given by
\begin{align}\label{eqn:njl_iso_density}
n_I = N_c \int_\Lambda \frac{d^3 k}{(2\pi)^3} \left( \frac{E_k + \frac{\mu_I}{2}}{E_k^+} - \frac{E_k - \frac{\mu_I}{2}}{E_k^-} \right) \, .
\end{align}
To find the physical pressure, we must take into account the considerations of Sec.~\ref{secII}. With $B^*$ defined in Eq.~\eqref{eqn:def_bstar}, the total pressure $P$ can be written as
\begin{align}
P = - \Omega_\textrm{NJL}(\sigma = \sigma_m; \Delta = \Delta_m)- B^*(\mu_I) \, ,
\end{align}
where $B^*$ only contributes in the case of a density dependent coupling $G = G(\mu_I)$. The energy density $\varepsilon$ is then defined as
\begin{align}
\varepsilon = - P + \mu_I n_I \, ,
\end{align}
and finally, the speed of sound squared $c_s^2$ is given by
\begin{align}
c_s^2 = \frac{\partial P}{\partial \varepsilon} \, .
\end{align}
\par A complete description of the model depends on the choice of three independent parameters: the current quark mass $m$, the ultraviolet momentum cutoff $\Lambda$ and the scalar/pseudoscalar coupling $G$. To avoid a possible misunderstanding, we denote $G = G_0$ for the fixed coupling case, and $G = G(\mu_I)$ for the case where it is calculated via fitting the model to the lattice data from Refs.~\cite{Brandt:2022fij,Brandt:2022hwy}. Finally, we adopt the parameter set $m = 4.76~\textrm{MeV}$, $\Lambda = 659~\textrm{MeV}$ and $G_0 = 4.78~\textrm{GeV}^{-2}$, which corresponds to a pion mass $m_\pi = 131.7~\textrm{MeV}$, pion decay constant $f_\pi = 92.4~\textrm{MeV}$ and quark condensate $\left\langle \bar{\psi}\psi \right\rangle^{1/3} = - 250~\textrm{MeV}$~\cite{Avancini:2019ego}.
%
%
\begin{figure}[t]
\centering
\includegraphics[width=\linewidth]{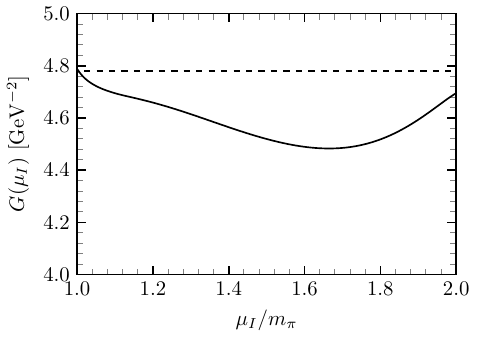}
\caption{Density dependent coupling $G(\mu_I)$ as a function of the normalized isospin chemical potential $\mu_I/m_\pi$}
\label{fig:g_of_mui_njl}
\end{figure}
\par Figure~\ref{fig:g_of_mui_njl} presents the behavior found for the isospin chemical potential dependent coupling $G(\mu_I)$ with the normalized isospin chemical potential $\mu_I/m_\pi$. The coupling starts as a decreasing function at $\mu_I = m_\pi$, until reaching a minimum around $\mu_I \approx 1.66 m_\pi$ and increasing thereafter, thus characterizing a non-monotonic behavior. The dashed line indicates the value of $G_0$ for the chosen set of parameters. It can be seen that the values of $G_0$ and $G(\mu_I)$ are the same at the starting value $\mu_I = m_\pi$ and split thereafter.
%
\begin{figure}[b]
\centering
\includegraphics[width=\linewidth]{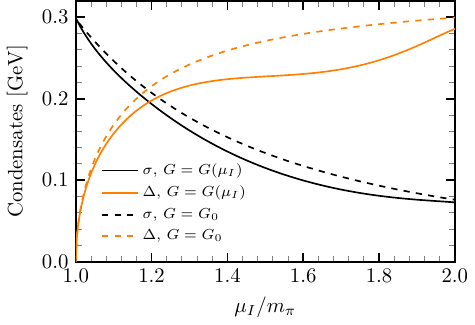}
\caption{Scalar condensate $\sigma$ and pion condensate $\Delta$ as functions of the normalized isospin chemical potential $\mu_I/m_\pi$, considering the cases of fixed coupling $G_0$ and density dependent coupling $G(\mu_I)$}
\label{fig:condensates_njl}
\end{figure}
\par In Fig.~\ref{fig:condensates_njl}, we show the chiral condensate $\sigma$ and pion condensate $\Delta$ as functions of the normalized isospin chemical potential $\mu_I/m_\pi$. For comparison both cases, the one with a fixed coupling $G_0$ and the one with a density dependent coupling $G(\mu_I)$, are presented. Both condensates have slightly lower values in the $G = G(\mu_I)$ case, and the pion condensate $\Delta$ shows some change in its curvature throughout the region of interest.
%
\begin{figure}[t]
\centering
\includegraphics[width=\linewidth]{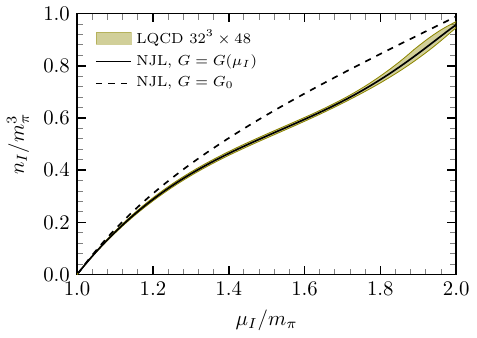}
\caption{Normalized isospin density $n_I / m_\pi^3$ as a function of the normalized isospin chemical potential $\mu_I/m_\pi$, considering the cases of fixed coupling $G_0$ and density dependent coupling $G(\mu_I)$}
\label{fig:density_njl}
\end{figure}
\par Figure~\ref{fig:density_njl} shows the behavior of the normalized isopin density $n_I/m_\pi^3$ as a function of the normalized isospin chemical potential $\mu_I / m_\pi$, considering the cases of fixed coupling $G_0$ and density dependent coupling $G(\mu_I)$. Also shown here are the LQCD results obtained in Refs.~\cite{Brandt:2022fij,Brandt:2022hwy} for a lattice size of $32^3 \times 48$, to which our density fit was performed. Consequently, the NJL model with $G(\mu_I)$ matches the LQCD behavior in this regime, while the fixed $G_0$ case does not reproduce the observed changes of slope.
%
\begin{figure}[b]
\centering
\includegraphics[width=\linewidth]{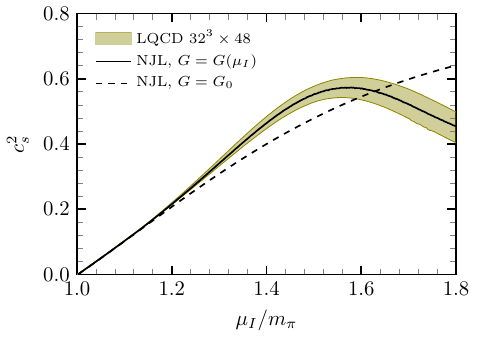}
\caption{Speed of sound $c_s^2$ as a function of the normalized isospin chemical potential $\mu_I/m_\pi$, considering the cases of fixed coupling $G_0$ and density dependent coupling $G(\mu_I)$}
\label{fig:cs2_njl}
\end{figure}
\par Finally, in Fig.~\ref{fig:cs2_njl} the square of the speed of sound $c_s^2$ is shown as a function of the normalized isospin chemical potential $\mu_I/m_\pi$ for both couplings $G_0$ and $G(\mu_I)$, together with the LQCD band for lattice size $32^3 \times 48$ obtained from Refs.~\cite{Brandt:2022fij,Brandt:2022hwy}. Again, the fixed coupling case does not reproduce the non-monotonic behavior found in the lattice results. The observed peak is only present in the NJL model if we consider the density dependent coupling $G(\mu_I)$ obtained by fitting the isospin density. Notice that the predicted behavior of the coupling can be traced back to that of the isospin density used for the fitting procedure. From Fig.~\ref{fig:density_njl}, we observe that the NJL model with a fixed coupling already provides a good description of the LQCD results at low $\mu_I$. The model curve with constant $G$ separates from the LQCD curve from above at intermediate values of $\mu_I$ to then come closer again to the LQCD curve for large values of $\mu_I$. This is reflected on the found properties of the coupling $G(\mu_I)$ in Fig.~\ref{fig:g_of_mui_njl}, as it decreases from the nominal value $G_0$ for low values of $\mu_I$, reaches a minimum around the inflection point observed in $n_I$ at intermediate values of $\mu_I$, and strengthens back again as the isospin density grows closer to the constant coupling case at high $\mu_I$. We also emphasize that the numerical changes in $G$ do not deviate too much from the nominal value $G_0$. Its maximum value is $\sim$ 4.8 GeV$^{-2}$ and its minimum is $\sim$ 4.5 GeV$^{-2}$ throughout the considered region, which corresponds to an overall variation of about $\sim 6\%$. Nonetheless, this dependence of $G$ with $\mu_I$ is directly linked to the non-monotonic behavior of the speed of sound, as the same good description found for $n_I$ reflects in the pressure and energy density due to thermodynamic consistency.
%
\section{The LSMq at finite isospin chemical potential}\label{secIV}

The LSMq is an effective theory that captures the approximate chiral symmetry of QCD. It describes the interactions among low-mass mesons and constituent quarks. In the following, we summarize the findings of Ref.~\cite{Ayala:2023cnt}, where the LSMq is used to describe the case of systems with an isospin charge imbalance.  

The LSMq Lagrangian is invariant under $SU(2)_{L}\times SU(2)_{R}$ chiral transformations
\begin{eqnarray}
\mathcal{L}&=&\frac{1}{2}(\partial_{\mu}\sigma)^{2}+\frac{1}{2}(\partial_{\mu}\vec{\pi})^{2}+\frac{a^{2}}{2}(\sigma^{2}+\vec{\pi}^{2})\nonumber\\
&-&\frac{\lambda}{4}(\sigma^{2}
+\vec{\pi}^{2})^{2} + i\bar{\psi}\gamma^{\mu}\partial_{\mu}\psi-ig\bar{\psi}\gamma^{5}\vec{\tau}\cdot \vec{\pi}\psi \nonumber\\
&-&g\bar{\psi}\psi\sigma,
\label{lsmqlag1}
\end{eqnarray}
where as before
$\vec{\tau}$
are the Pauli matrices, \begin{eqnarray}
\psi_{L,R}= \begin{pmatrix} u \\ d \end{pmatrix}_{L,R},
\label{doublet}
\end{eqnarray}is a  $SU(2)_{L,R}$ doublet, $\sigma$ is a real scalar singlet and $\vec{\pi}=(\pi_{1},\pi_{2},\pi_{3})$ is a triplet of real scalar fields. $\pi_3$ corresponds to the neutral pion whereas the charged ones are represented by the combinations
\begin{eqnarray}
    \pi_{\pm}=\frac{1}{\sqrt{2}}(\pi_{1} \mp i\pi_{2}).
    \label{pi1pi2basis}
\end{eqnarray}
The parameters $a^2$, $\lambda$ and $g$ are real and positive definite. 

A conserved isospin charge can be added to the LSMq Hamiltonian, multiplied by the isospin chemical potential $\mu_I$. The result is that the Lagrangian gets modified such that the ordinary derivative becomes a covariant derivative~\cite{Mannarelli:2019hgn}
\begin{eqnarray}
    \partial_{\mu} \to D_{\mu}= \partial_{\mu}+i\mu_{I} \delta_{\mu}^{0}, \nonumber \\\partial^{\mu} \to D^{\mu}= \partial^{\mu}-i\mu_{I} \delta_{0}^{\mu}.
\end{eqnarray}
Introducing also the spontaneous breaking of the chiral symmetry, the $\sigma$ field acquires a non-vanishing vacuum expectation value
\begin{equation*}
   \sigma \rightarrow \sigma+v.
\end{equation*}
To make better contact with the meson vacuum properties and to include a finite vacuum pion mass, $m_\pi$, we can add an explicit symmetry breaking term in the Lagrangian such that
\begin{equation}
    \mathcal{L} \to \mathcal{L'}=\mathcal{L} +h(\sigma + v).
    \label{lsmqlagwesb}
\end{equation}
The constant $h$ is fixed by requiring that the model expression for the neutral vacuum pion mass squared in the non-condensed phase corresponds to $m_\pi^2$.

In the non-condensed phase, the tree-level potential is
\begin{eqnarray}
V_{\text{tree}}=–\frac{a^2}{2} v^2 + \frac{\lambda}{4} v^4  - h v.
\label{Vtree}
\end{eqnarray}
The condensate $v_0$ is obtained from
\begin{eqnarray}
\frac{dV_{\text{tree}}}{dv} = (\lambda v^3\ –\ a^2 v -h)_{v=v_0}=0,
\label{dVtree}
\end{eqnarray}
or
\begin{eqnarray}
v_0(\lambda v_0^2\ –\ a^2)=h
\label{ordVtree}
\end{eqnarray}
The quantity in between parenthesis is precisely the square of the vacuum pion mass, $m_\pi^2$. Therefore
\begin{eqnarray}
h=m_\pi^2 v_0
\label{hident}
\end{eqnarray}
Also, notice that 
\begin{equation}
a^2 + m_\pi^2 = a^2 +\lambda v_0^2\ –\ a^2 = \lambda v_0^2,
\label{alsonotice}
\end{equation}
this yields
\begin{eqnarray}
h=m_\pi^2\sqrt{\frac{a^2+ m_\pi^2}{\lambda}}= m_\pi^2f_\pi,
\label{expressionforh}
\end{eqnarray}
where in the second equality we have used the Partially Conserved Axial Current (PCAC) statement to identify $m_\pi^2f_\pi$ with the symmetry breaking term represented in the LSMq by $h$, with $f_\pi$ being the pion decay constant. 

The introduction of $\mu_I$ further modifies the structure of the Lagrangian allowing for the emergence of meson condensates. Since we are interested in the dynamics of the pion fields, as a simplification in the pseudoscalar channels we work with the ansatz that $\langle \bar\psi i\gamma_5\tau_3\psi\rangle=0$ combined with $\langle \bar{u}i\gamma_5 d\rangle = \langle \bar{d}i\gamma_5 u\rangle^* \neq 0$~\cite{Mu:2010zz}. This corresponds to the existence of a Bose-Einstein condensate of the charged pions. Then, the charged pion fields can be referred from their condensates as 
\begin{equation}
\pi_{+}\to \pi_{+}+\frac{\Delta}{\sqrt{2}} e^{i\theta}, \quad \pi_{-}\to \pi_{-}+\frac{\Delta}{\sqrt{2}} e^{-i\theta},
\end{equation}
where the phase factor $\theta$ indicates the direction of the $U(1)_{I_3}$ symmetry breaking. We take $\theta=\pi$ for definitiveness. 

In the condensed phase $(\Delta\neq 0)$ the tree-level potential can be written as 
\begin{eqnarray}
V_{\rm tree}&=&-\frac{a^2}{2}\left(v^{2}+\Delta^2\right)+\frac{\lambda}{4}\left(v^2+\Delta^2 \right)^2 \nonumber \\ &-&\frac{1}{2}\mu_I^2\Delta^2-hv.
\label{treeeffectivepotential}
\end{eqnarray}

The fermion contribution to the one-loop effective potential becomes
\begin{equation}
\sum_{f=u,d}V_{f}^1= -2N_c\int\frac{d^3k}{(2\pi)^3}\left[E_\Delta^u +E_\Delta^d\right],
\label{fromfermions}
\end{equation}
with
\begin{eqnarray}
E_\Delta^u &=& \left\{\left(\sqrt{k^2+m_f^2}+\mu_I\right)^2+g^2\Delta^2\right\}^{1/2}\!\!\!, \nonumber \\
E_\Delta^d &=& \left\{\left(\sqrt{k^2+m_f^2}-\mu_I\right)^2+g^2\Delta^2\right\}^{1/2}\!\!\!, 
\label{fermionenergies}
\end{eqnarray}
where we chose that
\begin{eqnarray}
\mu_d&=&\mu_I\nonumber\\
\mu_u&=&-\mu_I.
\label{choice}
\end{eqnarray}
Notice that this choice differs from the one in Eq.~(\ref{choice1}).
Equation~(\ref{fromfermions}) is ultraviolet divergent. To identify the divergent terms, we work in the approximation whereby the fermion energies, Eqs.~(\ref{fermionenergies}), are expanded in powers of $\mu_I^2$
\begin{eqnarray}
\sum_{f=u,d}V_{f}^1&=& -2N_c\int\frac{d^3k}{(2\pi)^3}\Big( 2\sqrt{k^2+m_f^2+g^2\Delta^2}\nonumber\\
&+&\frac{\mu_I^2g^2\Delta^2}{(k^2+m_f^2+g^2 \Delta^2)^{3/2}}\Big).
\label{V1finf}
\end{eqnarray}

This expression can be readily computed using dimensional regularization in the $\overline{\mbox{MS}}$ scheme, with the result 
\begin{eqnarray}
\sum_{f=u,d}V_{f}^1&=&2N_c\frac{g^4\left(v^2+\Delta^2\right)^2}{(4\pi)^2}\left[\frac{1}{\epsilon}+\frac{3}{2} \right.\nonumber\\
&+&\left.\ln\left(\frac{\Lambda^2/g^2}{v^2+\Delta^2}\right)\right] -2N_c\frac{g^2\mu_I^2\Delta^2}{(4\pi)^2}\left[\frac{1}{\epsilon}\right.\nonumber\\
&+&\left.\ln\left(\frac{\Lambda^2/g^2}{v^2+\Delta^2}\right)\right],
\label{explVdiv}
\end{eqnarray}
where $N_c=3$ is the number of colors, $\Lambda$ is the dimensional regularization ultraviolet scale and the limit $\epsilon\to 0$ is to be understood.

To carry out the one-loop order renormalization of the effective potential, we introduce counter-terms that respect the structure of the tree-level potential and determine them by accounting for the stability conditions. These conditions require that the position of the minimum
in the $v$- and $\Delta$-directions remain the same as the tree-level potential ones. As a result a shift of the onset of pion condensation happens when the coupled
equations that determine the condensates receive loop corrections. Notice that this approach is different from the one followed in Ref.~\cite{Herpay:2008uw}, where the counter-terms are determined by requiring the finiteness of the propagator and the four-point boson vertex, or from the one followed in Ref.\cite{Chiba:2023ftg} where the renormalization is carried out using the expression for the Lagrangian in the non-condensed phase, introducing mass and coupling constants counter-terms. 

The tree-level minima in the $v$, $\Delta$ plane are found from
\begin{subequations}
    \begin{eqnarray}
\left.\frac{\partial V_{\rm tree}}{\partial v}=\left[\lambda v^3 -(a^2 - \lambda\Delta^2)v-h\right]\right|_{v_0,\,\Delta_0}&=&0 , \nonumber \\ \\
\left.\frac{\partial V_{\rm tree}}{\partial \Delta}=\Delta\left[\lambda\Delta^2-(\mu_I^2-\lambda v^2 + a^2)\right]\right|_{v_0,\,\Delta_0}&=&0 . \nonumber \\
\end{eqnarray}
\label{minima}
\end{subequations}
Notice that the second of Eqs.~(\ref{minima}) admits a real, non-vanishing solution, only when 
\begin{eqnarray}
\mu_I^2 > \lambda v^2 - a^2 = m_\pi^2,
\label{conditionmu}
\end{eqnarray}
which means that a non-zero isospin condensate is developed only when, for positive values of the isospin chemical potential, the latter is larger than the vacuum pion mass. This is what we identify as the condensed phase. The simultaneous solutions of Eqs.~(\ref{minima}) are
\begin{subequations}
   \begin{eqnarray}
v_0&=&\frac{h}{\mu_I^2},\\
\Delta_0&=&\sqrt{\frac{\mu_I^2}{\lambda} - \frac{h^2}{\mu_I^4} + \frac{a^2}{\lambda}}.
\end{eqnarray} 
\label{simultaneous}
\end{subequations}
Hereafter, we refer to the expressions in Eq.~(\ref{simultaneous}) as the classical solution.
\begin{figure}[t]
\centering
\includegraphics[width=\linewidth]{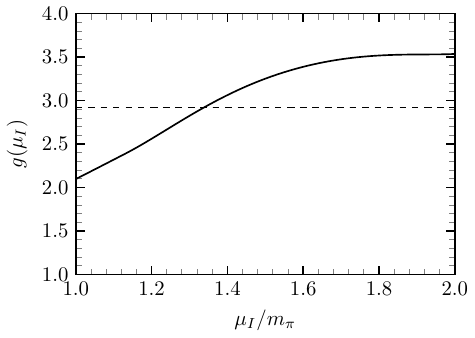}
\caption{Density dependent coupling $g(\mu_I)$ of the LSMq as a function of the normalized isospin chemical potential $\mu_I/m_\pi$. For comparison we also show the case of a constant coupling $g$ with the dashed line}
\label{fig:geffLSMq}
\end{figure}
The effective potential, up to one-loop order in the fermion fluctuations, including the counter-terms, can be written as
\begin{eqnarray}
V_{\rm eff}&=&V_{\rm tree}+\sum_{f=u,d}V_{f}^1-\frac{\delta\lambda}{4}(v^2+\Delta^2)^2\nonumber\\
&+&\frac{\delta a}{2}(v^2+\Delta^2)+\frac{\delta}{2}\Delta^2\mu_I^2.
\label{withcounterterms}
\end{eqnarray}
The counter-terms $\delta\lambda$ and $\delta$ are determined from the {\it gap equations} 
\begin{subequations}
    \begin{eqnarray}
\left.\frac{\partial V_{\rm eff}}{\partial v}\right|_{v_0,\,\Delta_0}&=&0,\\
\left.\frac{\partial V_{\rm eff}}{\partial \Delta}\right|_{v_0,\,\Delta_0}&=&0.
\end{eqnarray}
\label{stabilitycond}
\end{subequations}
These conditions suffice to absorb the infinities of Eq.~(\ref{explVdiv}). The counter-term $\delta a$ is determined by requiring that the slope of $V_{\rm eff}$ vanishes at $\mu_I=m_\pi$, 
\begin{eqnarray}
\left.\frac{\partial V_{\rm eff}}{\partial \mu_I}\right|_{\mu_I=m_\pi}=0,
\label{thirdcond}
\end{eqnarray}
or in other words, that the transition from the non-condensed to the condensed phase be smooth. The resulting effective potential is also $\Lambda$-independent. This can be seen by noticing that the coefficients of the $1/\epsilon$ terms are common to those of the $\ln(\Lambda^2)$ terms. Since the counter-terms cancel the $1/\epsilon$ divergence, they also  cancel the $\ln(\Lambda^2)$ dependence. Notice that by means of this procedure, the countertems are $\mu_I$-dependent. An alternative formulation whereby the counter-terms are found in the isospin symmetric phase, and thus considering that the broken chiral symmetry phase acts as the vacuum for the former, is currently being explored and will be reported elsewhere~\cite{AFP}.
\begin{figure}[!t]
\centering
\includegraphics[width=\linewidth]{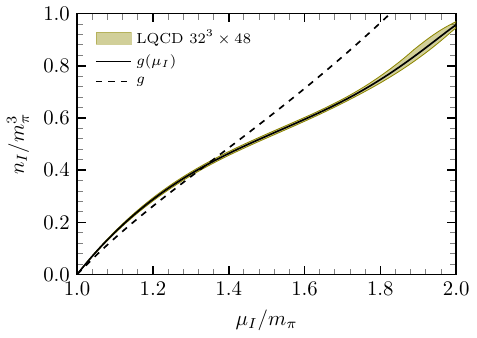}
\caption{Normalized isospin density $n_I / m_\pi^3$ as a function of the normalized isospin chemical potential $\mu_I/m_\pi$, considering the cases of LSMq fixed coupling $g$ and density dependent coupling $g(\mu_I)$}
\label{fig:isodensitylsmq}
\end{figure}

The model requires fixing three independent parameters: the boson self-coupling $\lambda$, the boson-fermion coupling $g$ and the mass parameter $a$. For this purpose, notice that in vacuum we have
\begin{equation}
m_\sigma^2\ –\ 3m_\pi^2 = 3\lambda v_0^2\ –\ a^2\ –\ 3\lambda v_0^2 + 3a^2 = 2a^2,
\end{equation}
or
\begin{equation}
a=\sqrt{\frac{m_\sigma^2\ –\ 3m_\pi^2}{2}}.
\end{equation}
Also 
\begin{equation}
g = \frac{m_q}{v_0} = \frac{m_q}{f_\pi},
\end{equation}
and $\lambda$ is obtained from Eq.~(\ref{expressionforh}) as
\begin{equation}
\lambda=\frac{m_\sigma^2-m_\pi^2}{2f_\pi^2}.
\end{equation}
For $m_q=270$ MeV, $m_\sigma=450$ MeV, $m_\pi=131.7$ MeV and $f_\pi=92.4$ MeV, one readily obtains $\lambda=10.84$, $a=274.29$ MeV and $g=2.9$. These values are chosen to provide a reasonably good description of the LQCD data, particularly for the pressure~\cite{Brandt:2022hwy,Brandt:2022fij}. In what follows, we concentrate on finding the $\mu_I$ induced correction to $g$ treating $\lambda$ as not being influenced by the medium. The reason for this choice is that in our approach, fermions induce loop effects whereas bosons are treated as average fields. Thus, the coupling that involves fermions, namely $g$, should be treated as the only coupling receiving medium modifications.

\begin{figure}[t!]
\centering
\includegraphics[width=\linewidth]{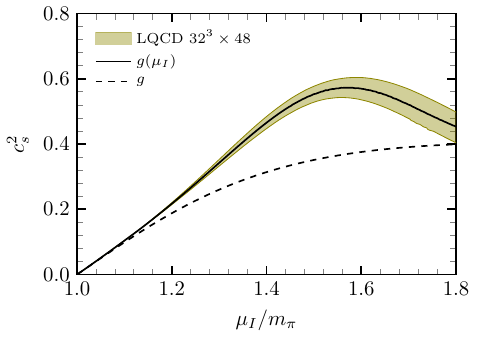}
\caption{Square of the speed of sound $c_s^2$ as a function of the normalized isospin chemical potential $\mu_I/m_\pi$ using the LSMq density dependent coupling $g(\mu_I)$. For comparison, shown is also $c_s^2$ using the constant value for $g$}
\label{fig:cs2_lsmq}
\end{figure}

Figure~\ref{fig:geffLSMq} shows the isospin chemical potential dependent effective coupling $g(\mu_I)$ as a function of the normalized isospin chemical potential $\mu_I/m_\pi$ obtained by fitting the LSMq expression for the isospin density to the LQCD data from Ref.~\cite{Brandt:2022fij,Brandt:2022hwy}.
Figure~\ref{fig:isodensitylsmq} shows the LSMq normalized isospin density as a function of the normalized isospin chemical potential $\mu_I/m_\pi$ using the obtained $g(\mu_I)$ from the LQCD data from Ref.~\cite{Brandt:2022fij,Brandt:2022hwy}. For comparison, we also show the LSMq result using a constant coupling $g$.
Figure~\ref{fig:cs2_lsmq} shows the LSMq square of the speed of sound $c_s^2$ as a function of the normalized isospin chemical potential $\mu_I/m_\pi$ using the density dependent coupling $g(\mu_I)$, compared to the LQCD data from Ref.~\cite{Brandt:2022fij,Brandt:2022hwy}. For comparison, we also show the LSMq result using a constant coupling $g$. Notice that the peak for $\mu_I\sim 1.5\ m_\pi$ is well reproduced by employing the found profile of $g(\mu_I)$.

We notice from Fig.~\ref{fig:geffLSMq} that, as opposed to the result for the NJL model, $g(\mu_I)$ is a monotonically increasing function of $\mu_I$. The figure also shows that the constant value (represented by the dashed line) corresponds to the average of $g(\mu_I)$ in the considered range for the analysis. This behavior can be understood recalling that a constant value can only describe the gross features of the isospin density dependence on $\mu_I$. In order to describe the detailed features, it can then be expected that for some range of $\mu_I$ the $g$-values are below this average whereas for another range the $g$-values are above, so that on average $g$ takes its given constant value. Here, $g(\mu_I)$ starts below the constant value because it is computed from the properties of the isospin density, which in turn is computed as the derivative of the pressure with respect to $\mu_I$. According to the results of Ref.~\cite{Ayala:2023cnt}, the latter starts off from $\mu_I=m_\pi$ with a smaller value of the derivative as compared with the LQCD data. It may also be noted that the value of $g(\mu_I/m_\pi=1)$ does not coincide with that of the NJL model coupling $G(\mu_I/m_\pi=1)$, first because the two constants do not have the same units - $g$ is dimensionless whereas $G$ has units of inverse mass squared. It is nevertheless important to point out that if the coupling $G$ is multiplied by the square of the ultraviolet cutoff $\Lambda$, the other dimensionful quantity associated to vacuum properties, one obtains $G\Lambda^2\simeq 2.07$, which coincides with the value of $g(\mu_I/m_\pi=1)$. From a more detailed viewpoint, $g$ and $G$ are quantities that capture the physics of systems possessing chiral symmetry in a different manner: In the NJL model, as can be seen from Eq.~\eqref{eqn:njl_omega}, $G$ (or rather, its inverse) represents the strength with which the condensates contribute to the systems energy. In the LSMq, $g$ has a similar role, as can be seen from Eq.~\eqref{explVdiv}, but this behavior is obtained using in addition the explicit expression for the quark mass $m_q$ in terms of the coupling $g$ and the chiral condensate $v$.

%
\section{Summary and discussion}\label{concl}
Recall that many of the results from LQCD simulations and/or universality arguments, such as the chiral transition in QCD at nonzero temperature and baryon densities, can be successfully reproduced using effective models. It is therefore intriguing that such models have difficulties to reproduce, even at a qualitative level, the peak of the speed of sound as a function of $\mu_I$, recently reported by LQCD~\cite{Brandt:2022hwy}. Particularly, calculations based on NJL type of models~\cite{Avancini:2019ego,Lopes:2021tro}, the LSMq and even chiral perturbation therory~\cite{Ayala:2023cnt} have not found this peak in $c_s^2$.

Our results show that an interesting way of reconciling results for the non-monotonic behavior of $c_s^2$ in isospin imbalanced strongly interacting matter, obtained with effective models and LQCD, may be closely connected with the introduction of medium-dependent couplings. In this work, we have determined the dependence of the couplings with $\mu_I$ using as input the three-flavor LQCD data for the isospin density $n_I$~\cite{Brandt:2022hwy}, and the two-flavor version of the NJL and LSMq. We verified that the introduction of a medium dependence through the coupling requires a careful analysis of the thermodynamics of the system, and the addition of a compensating function can recover the agreement between statistical mechanics definitions and thermodynamic relations~\cite{Gorenstein:1995vm}. As the isospin density was fitted to the lattice, the numerical results for the thermodynamic functions are consequently also in agreement with LQCD results. Again, our idea here is not to claim a perfect description, but to point out that medium contributions to the coupling, which are normally not accounted for in effective models, may be closely connected to the characterization of a non-monotonic behavior in the speed of sound. We have also argued that the use of two-flavor models to compare to a three-flavor LQCD result is justified since the effect of the strange quark on $\langle n_I\rangle$ is, in our effective model description, only captured at higher orders which are suppressed with respect to the leading order calculation shown in this work. Nevertheless, it is important to mention that we have results from a preliminary analysis, using the the three-flavor NJL model of Ref.~\cite{Lopes:2021tro}, to explore whether the description of this effect is different from what we obtain with the two-flavor models. What we find is that the results for the $G$-coupling are very similar in the two cases, showing the same overall shape as a function of $\mu_I$, decreasing towards a minimum and increasing afterwards. The slightly different values can then be attributed to the effect of strangeness, verifying that this  has quantitatively a small effect. The similarities reflect the fact that, although the increase of degrees of freedom in the effective model may correspond to a fairer description of LQCD data, the two-flavor models already capture the physics encoded in the behavior of the coupling that is in turn responsible for the peak in the speed of sound. This result is part of an ongoing effort and will be complemented with a corresponding three-flavor analysis in the LSMq, which is more involved and will be reported elsewhere.

We find that the $\mu_I$ dependence of the couplings in the NJL and LSMq models turn out to have different behaviors. However, notice also that as opposed to the former, in the latter case we have only considered a medium-induced modification of $g$, ignoring possible medium-induced modifications of $\lambda$. The rationale behind this approximation rests on the assumption that since quantum effects are introduced in terms of a fermion loop~\cite{Ayala:2023cnt}, it is only the fermion coupling which should be considered to receive medium modifications. To try reconciling these behaviors, improvements to the approximations considered in LSMq calculations are currently being performed and will also be reported elsewhere.


\section*{Acknowledgments}
Support for this work was received in part by UNAM-PAPIIT IG100322 and by Consejo Nacional de Humanidades, Ciencia y Tecnolog\'ia grant numbers CF-2023-G-433, A1-S-7655 and A1-S-16215. This work was also partially supported by Conselho Nacional de Desenvolvimento Cient\'ifico 
e Tecno\-l\'o\-gico  (CNPq), Grants No. 309598/2020-6 (R.L.S.F.) and No. 141270/2023-3 (B.S.L.); 
Funda\c{c}\~ao de Amparo \`a Pesquisa do Estado do Rio 
Grande do Sul (FAPERGS), Grants Nos. 19/2551- 0000690-0 and 19/2551-0001948-3 (R.L.S.F.); The work is also part of the project Instituto Nacional de Ci\^encia 
e Tecnologia - F\'isica Nuclear e Aplica\c{c}\~oes (INCT - FNA), Grant No. 464898/2014-5.
\bibliography{cs2peak_bib}

\end{document}